\documentclass[num-refs]{wiley-article}

\usepackage{siunitx}
\usepackage{hanging}
\usepackage[english]{babel}  
\usepackage{hyphenat}

\papertype{Original Article}
\paperfield{Journal Section}

\title{Transforming Teachers’ Roles and Agencies in the Era of Generative AI: Perceptions, Acceptance, Knowledge, and Practices}

\author[1,2,3\authfn{1}]{Xiaoming Zhai}

\affil[1]{AI4STEM Education Center, University of Georgia, Athens, GA 30666}
\affil[2]{National GENIUS Center, University of Georgia, Athens, GA 30666}
\affil[3]{Department of Mathematics, Science, and Social Studies Education, University of Georgia, Athens, GA 30666}

\corraddress{125M Aderhold Hall, 110 Carlton St., Athens, GA 30602}
\corremail{xiaoming.zhai@uga.edu}

\fundinginfo{This study is supported by the National Science Foundation (\#2101104) and Institute of Education Sciences (\#R305C240010). Any opinions, findings, conclusions, or recommendations expressed in this material are those of the author(s) and do not necessarily reflect the views of the NSF or IES.}

\runningauthor{Xiaoming Zhai}

\begin{document}

\begin{frontmatter}
\maketitle

\begin{abstract}
This paper explores the transformative impact of Generative Artificial Intelligence (GenAI) on teachers’ roles and agencies in education, presenting a comprehensive framework that addresses teachers’ perceptions, knowledge, acceptance, and practices of GenAI. As GenAI technologies, such as ChatGPT, become increasingly integrated into educational settings, teachers are required to adapt to evolving classroom dynamics, where AI plays a significant role in content creation, personalized learning, and student engagement. However, existing literature often treats these factors in isolation, overlooking how they collectively influence teachers’ ability to effectively integrate GenAI into their pedagogical practices. This paper fills this gap by proposing a framework that categorizes teachers into four roles— Observer, Adopter, Collaborator, and Innovator— each representing different levels of GenAI engagement, outlining teachers’ agencies in GenAI classrooms. By highlighting the need for continuous professional development and institutional support, we demonstrate how teachers can evolve from basic GenAI users to co-creators of knowledge alongside GenAI systems. The findings emphasize that for GenAI to reach its full educational potential, teachers must not only accept and understand its capabilities but also integrate it deeply into their teaching strategies. This study contributes to the growing literature on GenAI in education, offering practical implications for supporting teachers in navigating the complexities of GenAI adoption.

\end{abstract}
\end{frontmatter}

\section{Introduction}
The integration of Generative Artificial Intelligence (GenAI) in educational environments is rapidly transforming the landscape of teaching and learning, presenting both unprecedented opportunities and significant challenges for educators. GenAI technologies, including advanced natural language processing (Zhai et al., 2020), content creation tools (Lee \& Zhai, 2024b), and interactive learning assistants (Latif et al., 2024; Pahi et al., 2024), offer the potential to personalize education, enhance student engagement, and revolutionize the production of educational materials. These advancements promise to disrupt traditional pedagogical practices, fundamentally altering the dynamics of classroom instruction and the relationships between teachers and students and technologies (Pahi et al., 2024; Tang et al., 2024). That is, as GenAI becomes more sophisticated and widespread, it is increasingly capable of performing tasks that were once the sole domain of teachers, such as generating customized learning content (Nyaaba, Shi, et al., 2024) and providing nuanced feedback (Zhai, 2023). This shift necessitates a re-examination of the roles teachers play within the classroom and their agencies-- the capacity and autonomy of teachers to make decisions, take initiative, and act in ways that shape their teaching practices and professional development, as they are now required to adapt to an environment where GenAI is not merely an auxiliary tool but an active collaborator in the instructional process (Lan \& Chen, 2024). Understanding how teachers perceive, accept, and engage with these emerging technologies is crucial for ensuring that GenAI is integrated in a way that enhances, rather than diminishes, the educational experience (Collie \& Martin, 2024).

Despite the rapidly growing presence of GenAI in education, the literature remains limited in its exploration of the nuanced roles that teachers assume as they interact with these systems. Existing research has primarily concentrated on the technical efficacy of GenAI tools and their perceived impacts on educational outcomes, often neglecting the human element—specifically, how teachers adapt to and collaborate with these technologies and how GenAI alters teachers’ agencies in classrooms (Kizilcec et al., 2024). Studies that do focus on teachers tend to examine their attitudes and perceptions towards GenAI adoption in general (Garofalo \& Farenga, 2024; Guo et al., 2024) without delving deeply into the specific challenges and opportunities presented by GenAI. Furthermore, the literature frequently assumes a one-size-fits-all approach to technology adoption, failing to account for the varying levels of GenAI integration across different educational contexts and how these differences might influence teachers’ roles and agencies. Specifically, the focus has often been on the capabilities of GenAI tools rather than on the pedagogical implications these tools have for the relational dynamics in the classroom. Consequently, there is a critical need for research to examine the specific ways in which GenAI influences teachers’ roles and agencies, going beyond simplistic notions of technology adoption to explore the deeper, more complex interactions between educators and GenAI systems.

Moreover, current theoretical frameworks that describe teachers’ roles and agencies in relation to educational technology, such as the Technological Pedagogical Content Knowledge (TPACK) model (Rosenberg \& Koehler, 2015), may no longer fully capture the breadth of changes brought about by GenAI. These models often emphasize the enhancement of existing pedagogical practices through the integration of technology, focusing on how teachers can effectively incorporate digital tools into their instruction. However, they may fall short in addressing the transformative potential of GenAI, which not only supports but also actively shapes both short- and long-term educational practices through its ability to generate new content and facilitate adaptive learning environments (Mishra et al., 2024). The TPACK model, for instance, is primarily concerned with the intersection of technology, pedagogy, and content knowledge, and while it provides valuable insights into how teachers can integrate technology into their teaching, it does not fully account for the profound ways in which GenAI can alter the roles of educators (Celik, 2023; Mishra et al., 2023). GenAI’s ability to produce creative outputs, such as personalized lesson plans, assessment tasks and rubrics, and even artwork, introduces new dynamics into the classroom that go beyond the scope of traditional technology-enhanced learning (Feldman-Maggor et al., 2024). Furthermore, GenAI technologies have the potential to create entirely new pedagogical approaches, requiring teachers to move beyond traditional agencies and embrace more innovative and creative ways of facilitating learning. This includes a shift from being mere facilitators to becoming co-creators of knowledge alongside GenAI systems, a role that current frameworks do not fully address (Kong \& Yang, 2024). The lack of a comprehensive framework that acknowledges this transformation presents a significant research gap, particularly as GenAI continues to evolve and become more integral to educational settings.

To fill these gaps, this paper aims to present a conceptual framework that depicts the transformation of teachers’ roles and agencies in the era of GenAI. The proposed framework identifies four distinct roles that teachers may adopt—Observer, Adopter, Collaborator, and Innovator—each corresponding to different stages of GenAI integration in the classroom. By examining these roles and agencies through the lenses of teachers’ (including pre-service teachers) perception, acceptance, knowledge, and practice, as well as considering the nature of GenAI and the evolving relationships between teachers and these technologies, this paper offers a nuanced understanding of the transformative potential of GenAI in teaching and learning. The framework not only provides a theoretical foundation for future empirical research but also offers practical insights for educators and policymakers, seeking to navigate the complexities of GenAI integration in educational settings.
\section{Nature of Generative AI and Teachers’ Roles and Agencies}

The advent of GenAI marks a significant evolution in the capabilities of AI systems, characterized by their ability to autonomously produce content that was traditionally generated by humans. Unlike conventional technological systems, which are primarily designed to perform specific tasks based on predefined rules and datasets, GenAI leverages advanced machine learning models, such as deep neural networks, to create new content, ranging from text and images to audio and video (Jo, 2023). This capacity for content generation introduces a paradigm shift in the educational landscape, where the role of teachers is increasingly influenced by the interplay between human creativity and machine-generated outputs (Nyaaba \& Zhai, 2024). Understanding the nature of GenAI is critical for comprehending its potential impact on teachers’ roles and agencies, as it extends beyond mere automation to the co-creation and augmentation of educational materials and experiences (Pahi et al., 2024).

GenAI systems are distinguished by their ability to learn from vast datasets and generate new outputs that are not simply reproductions of existing data but rather novel creations that can adapt to different contexts and requirements. For example, in education, GenAI can produce customized learning materials tailored to individual student needs, such as personalized textbooks, assessments, and interactive simulations (Bewersdorff et al., 2024; van den Berg \& du Plessis, 2023). This capability challenges traditional teaching practices by shifting some aspects of content creation from the teacher to the GenAI systems. As a result, teachers’ roles in content delivery and curriculum design are transformed, with GenAI acting as both a collaborator and an enhancer of instructional content (Lee \& Zhai, 2024a). Teachers may no longer need to spend as much time on creating or sourcing instructional materials; instead, they can focus on curating, guiding, and refining AI-generated content to meet the specific learning objectives and needs of their students.

Furthermore, the nature of GenAI allows for the creation of dynamic and adaptive learning environments that can respond in real-time to student interactions. For instance, GenAI can generate instant feedback on student assignments, simulate complex scenarios for experiential learning, or even create virtual environments where students can engage in immersive educational experiences (Ali et al., 2023; Goldberg \& Robson, 2023). This capability not only enhances the learning process but also redefines the teachers’ roles from being the primary source of knowledge to becoming a facilitator of an interactive, AI-enhanced learning environment. In this context, teachers are required to develop new competencies in managing and integrating AI-generated content and experiences, ensuring that they align with pedagogical goals and ethical standards (Feldman-Maggor et al., 2024). This shift highlights the need for teachers to be not just users of GenAI tools, but also critical evaluators and integrators of AI outputs, guiding the learning process in ways that harness the strengths of both human and machine intelligence.

The introduction of GenAI also brings with it new challenges and ethical considerations that directly impact teachers’ roles and agencies. One such challenge is the potential for bias in AI-generated content, as GenAI systems are trained on large datasets that may reflect existing biases in the data (Srinivasan \& Uchino, 2021; Zhou et al., 2024). Teachers must be vigilant in identifying and addressing any biases that may emerge in AI-generated materials, ensuring that the content is equitable and inclusive (Zhai \& Nehm, 2023). Additionally, the use of GenAI in education raises questions about intellectual property and authorship, particularly when AI systems generate original content (Cotton et al., 2024). Teachers need to navigate these complexities, balancing the use of AI-generated content with considerations of authorship, originality, and academic integrity.

Moreover, the integration of GenAI into educational practices prompts a rethinking of the teacher-student relationship. As GenAI systems take on more active roles in content creation and interaction with students, the traditional hierarchical dynamic between teacher and student may be altered (Shi \& Choi, 2024). Students might begin to view AI as an authoritative source of knowledge, potentially diminishing the teacher’s perceived authority and expertise (Tang \& Cooper, 2024). To mitigate this, teachers must redefine their roles and agencies to emphasize aspects of teaching that cannot be replicated by GenAI, such as providing emotional support, fostering critical thinking, and nurturing creativity. By focusing on these uniquely human elements of instruction, teachers can maintain their central role in the classroom, even as GenAI takes on a more prominent position in the learning processes.

Given the nature of GenAI, it is fundamentally altering the traditional roles of teachers, introducing new opportunities for content creation, personalized learning, and interactive educational experiences. These changes go beyond purely automation by incorporating intelligence and creativity in the instruction processes and learning outcomes. However, the challenges brought by GenAI are not ignorable, requiring teachers to adapt and expand their attitudes and skill sets. As GenAI continues to evolve, the roles and agencies of teachers will increasingly involve managing and integrating GenAI-generated content, addressing ethical concerns, and maintaining the human elements of teaching that are essential for student development. Understanding these dynamics is crucial for developing a comprehensive framework that supports teachers in navigating the complexities of GenAI in education.

\section{Teachers’ Perceptions, Acceptance, Knowledge, and Practices of Generative AI}

As GenAI continues to gain traction in educational settings, understanding teachers’ perceptions, acceptance, knowledge, and practices regarding this technology becomes increasingly crucial. GenAI is unlikely to replace teachers; instead, it plays a crucial role as an instructional companion to assist teachers in all aspects of instruction. Teachers’ perceptions of GenAI thus directly impact their acceptance, which, together with their knowledge of GenAI, further influences how effectively GenAI is integrated into the classroom and shapes the broader impact it has on teaching and learning. This section delves into the existing literature on these aspects, highlighting how teachers’ attitudes and knowledge affect their engagement with GenAI and, consequently, the educational outcomes of its application.

\subsection{Perceptions of Generative AI}
Teachers’ perceptions of GenAI are shaped by a combination of familiarity with the technology, understanding of its capabilities, and beliefs about its potential benefits and risks. Research suggests that teachers who perceive GenAI as a tool that can enhance their instructional practices are more likely to embrace its uses in the classroom (Kaplan-Rakowski et al., 2023). Positive perceptions often stem from recognizing GenAI’s ability to personalize learning, reduce administrative burdens, and provide innovative teaching materials. Specifically, Collie and Martin (2024) found that teachers’ perceived autonomy-supportive leadership, professional growth striving, and change-related stress of GenAI are significantly associated with their eventual engagement and use of GenAI. However, there are also significant concerns. Teachers may perceive GenAI as a threat to their professional roles, fearing that the technology could replace certain teaching functions or diminish the need for teachers. Additionally, there is apprehension about the ethical implications of AI-generated content, such as the potential for bias, pseudo bias, the loss of originality, and issues related to data privacy and security (Zhai \& Krajcik, 2024). These concerns can lead to resistance or skepticism towards the adoption of GenAI, particularly among educators who are less familiar with the technology or who have had limited exposure to its applications.

\subsection{Acceptance of Generative AI}
Acceptance of GenAI in educational contexts is influenced by several factors, including perceived ease of use, perceived usefulness, and the alignment of the technology with teachers’ pedagogical goals. The technology acceptance model  (TAM) and the unified theory of acceptance and use of technology (UTAUT) provide valuable frameworks for understanding these dynamics (Davis et al., 1989; Venkatesh et al., 2016). According to these models, teachers are more likely to accept and use GenAI if they believe it will enhance their teaching effectiveness and if the technology is user-friendly. However, the acceptance of GenAI also hinges on the extent to which teachers feel confident in their ability to control and integrate technology into their pedagogical practices. This confidence is often tied to the availability of professional development and training opportunities that equip teachers with the necessary skills to use GenAI effectively (Nyaaba \& Zhai, 2024). Studies have shown that teachers’ sentiment of GenAI lacks homogeneity (D. Lee et al., 2024). That is, when teachers receive adequate support and training, their acceptance of GenAI increases, leading to more meaningful and sustained use of the technology in their classrooms.

\subsection{Knowledge of Generative AI}
Teachers’ knowledge of GenAI plays a critical role in determining how they perceive, accept, and ultimately utilize the technology. This knowledge encompasses both technical understanding and pedagogical integration. Teachers need to be familiar with the underlying mechanisms of GenAI, including how these systems learn, generate content, and potentially introduce bias. A deep understanding of these concepts enables teachers to critically assess the outputs produced by GenAI and to make informed decisions about how to incorporate these outputs into their teaching. On the pedagogical side, knowledge of best practices for integrating GenAI into instructional design is essential. This includes understanding how to use GenAI-generated content to support differentiated instruction, foster student engagement, and enhance learning outcomes, which sometimes go beyond the traditional TPACK framework (Feldman-Maggor et al., 2024). Research indicates that teachers who possess a strong foundational knowledge of GenAI are better equipped to leverage its capabilities, resulting in more effective and innovative teaching practices (Garofalo \& Farenga, 2024). Conversely, a lack of knowledge can lead to superficial or inappropriate use of the technology, which may undermine its potential benefits.

\subsection{Practices Involving Generative AI}
The ways in which teachers incorporate GenAI into their teaching practices vary widely and are influenced by their perceptions, acceptance, and knowledge of the technology. Some educators use GenAI to automate routine tasks, such as grading or creating instructional materials, freeing up time to focus on more complex and creative aspects of teaching (Laak \& Aru, 2024; Lee, Latif, et al., 2023; G.-G. Lee et al., 2024; Lee \& Zhai, 2023). Others integrate GenAI-generated content into their lessons, using it to create personalized learning experiences or to stimulate student curiosity and creativity. For example, teachers might use GenAI-generated prompts for writing assignments, or they might employ GenAI-driven materials to help students explore complex scientific concepts (Martin \& Graulich, 2024; Nyaaba, Shi, et al., 2024). However, the effectiveness of these practices is highly dependent on how well teachers understand and manage the capabilities of GenAI. Effective practices typically involve a thoughtful balance between GenAI-generated content and human oversight, ensuring that the technology enhances rather than detracts from the educational experience. Moreover, best practices in the use of GenAI also involve continuous reflection and adaptation, as teachers assess the impact of the technology on student learning and make adjustments to their instructional strategies accordingly.

As teachers’ perceptions, acceptance, knowledge, and practices regarding GenAI are interrelated factors that play central roles in teachers’ uses of GenAI, they collectively influence the integration of this technology into educational settings. Positive perceptions and high levels of acceptance are often associated with greater knowledge and more effective use of GenAI, while negative perceptions and low acceptance can hinder its adoption and impact (Yin et al., 2024). As GenAI continues to evolve, ongoing professional development and support will be crucial in helping teachers navigate this complex landscape, enabling them to harness the full potential of the technology to enhance teaching and learning. This understanding lays the groundwork for the section to follow, which will introduce a conceptual framework that captures the transformative roles of teachers in the era of GenAI.

\section{A Framework: Transforming Teachers’ Roles and Agencies in the Era of Generative AI}
The evolution of teachers’ roles and agencies in the context of GenAI can be understood as a dynamic and multifaceted process, driven by the interplay between the nature of GenAI technologies and teachers’ perceptions, acceptance, knowledge, and practices. These stages and teachers’ roles are deeply rooted in well-established models of technology adoption and innovation diffusion, such as the TAM and UTAUT (Davis et al., 1989; Venkatesh et al., 2016). These models emphasize that the adoption of new technologies follows a continuum, beginning with awareness and moving toward active engagement and leadership. In the context of GenAI, this continuum is particularly pronounced due to the technology’s complexity and the profound impact it has on teaching practices. This framework, which identifies four distinct stages—Awareness, Exploration, Integration, and Innovation—each corresponding to a specific role (i.e., GenAI Observer, Adaptor, Integrator, and Innovator), is not arbitrary but rather emerges from a careful analysis of how GenAI fundamentally alters the educational landscape and, consequently, the professional identity and practices of educators (see Figure 1).

GenAI differs significantly from previous educational technologies due to its capacity for autonomous content creation, adaptability, and integration into complex pedagogical processes. Unlike earlier forms of AI, GenAI introduces new possibilities for personalized learning, creative content generation, and real-time interaction with students. These capabilities demand a reconfiguration of traditional teaching roles, as educators must navigate not only the technical aspects of GenAI but also its implications for pedagogy, ethics, and classroom dynamics. The four stages of the proposed framework are identified based on how teachers progressively engage with these novel aspects of GenAI, moving from initial exposure to deep integration and innovation. The stages identified in this framework are informed by the observed progression in teachers’ interactions with GenAI, as documented in the literature: from passive observation and initial experimentation to full integration and ultimately, the pioneering of new pedagogical paradigms.
\begin{figure}[bt]
\centering
\includegraphics[width=8cm]{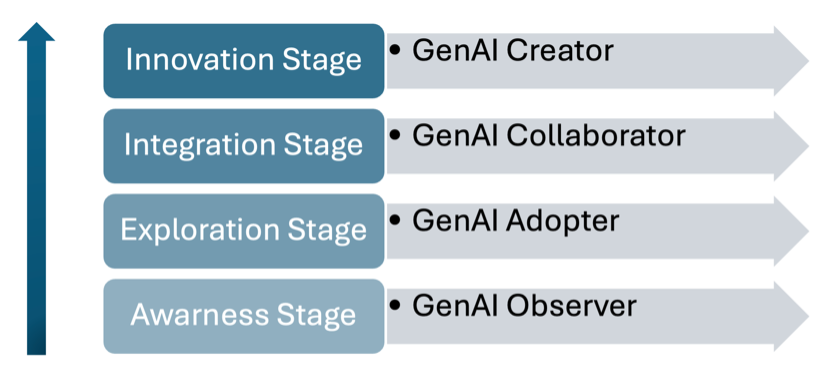}
\caption{A Framework: Transforming Teachers’ Roles and Agencies in the Era of Generative AI}
\end{figure}
\subsection{Awareness Stage: GenAI Observer}
At the Awareness Stage, teachers function primarily as observers of GenAI, beginning to gather information and form initial impressions about the technology’s role in education. This stage is characterized by a cautious approach, as teachers have yet to actively engage with AI tools in their teaching practices. Research suggests that teachers at this stage are often motivated by a need to understand the broader educational trends and potential impacts of GenAI on the classroom or pure curiosity about GenAI tools such as ChatGPT (Kizilcec et al., 2024). They are in the early phases of exposure, relying heavily on social media advertisements and knowledge acquired from colleagues, professional development sessions, educational conferences, or academic readings.

As a result, the practice of GenAI in classroom settings at this stage is limited or nonexistent. Teachers might explore simple GenAI tools, such as ChatGPT, for learning purposes, but these interactions are sporadic and exploratory rather than integrated into their regular teaching routines. For example, in their study, Nyaaba, Shi, et al. (2024) found that teachers explored GenAI in their own learning as a learning buddy. That is, teachers are more likely to observe the classroom uses of GenAI applications from a distance, considering their potential without fully committing to their uses. As shown in Figure 2, both teachers showed strong interest in leveraging GenAI for classroom teaching.

Teachers at this stage usually possess a basic awareness of GenAI and its potential implications for education. They recognize that GenAI is becoming increasingly relevant to their profession, but their understanding is largely superficial, centered around general concepts rather than specific applications. This nascent awareness is often accompanied by a mixture of curiosity and apprehension (e.g., potential bias, the possibility to replace teachers) as teachers grapple with the dual possibilities of GenAI either enhancing or disrupting their established teaching methods. Meanwhile, teachers’ knowledge of GenAI is minimal and predominantly theoretical. They may have encountered basic concepts related to GenAI, such as machine learning and natural language processing, but this knowledge is not yet connected to practical applications within the classroom. Their understanding of GenAI is often abstract, focused more on what GenAI could theoretically do rather than how it could be used effectively in educational settings. Some teachers may even have a misunderstanding of AI concepts or AI bias, as other people would have done (Bewersdorff et al., 2023; Zhai \& Krajcik, 2024), which can impact their further exploration of GenAI for classroom uses.

Due to the knowledge and practice gaps, teachers’ acceptance of GenAI at this stage is typically low. Teachers may be skeptical of the technology’s purported benefits, particularly if they perceive GenAI as a threat to their professional autonomy or job security. Research has frequently highlighted such skepticism, noting that resistance to technology such as GenAI integration often stems from a lack of familiarity and confidence in using the technology (Lee \& Song, 2024). Teachers might also harbor concerns about the ethical implications of GenAI, including issues related to data privacy, student autonomy, and the potential for GenAI to perpetuate existing biases.
\begin{figure}[bt]
\centering
\includegraphics[width=12cm]{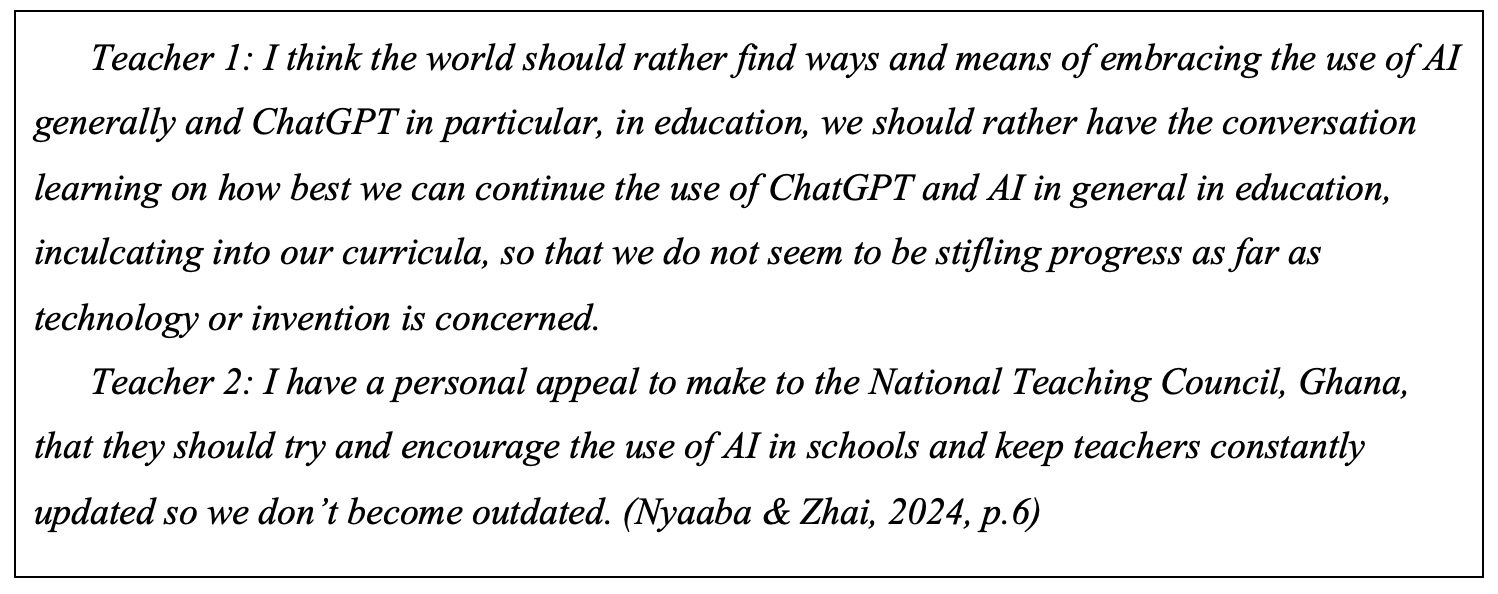}
\caption{Quotations from teachers as GenAI Observers, who showed interest in GenAI}
\end{figure}

\subsection{Exploration Stage: GenAI Adopter}
As teachers progress to the Exploration stage, they begin to engage more actively with GenAI technologies, piloting their applications in the classroom. This stage marks a transition from passive observation to active exploration, where teachers start to integrate GenAI tools into their teaching practices, albeit in a trial-and-error manner. Teachers may test the waters of GenAI integration, seeking to understand its practical benefits and limitations.

Practice at this stage is characterized by exploration and piloting. Teachers use GenAI-driven tools such as ChatGPT, intelligent tutoring systems, automated grading software, and GenAI-based learning platforms to supplement their lessons. For example, in their study, Nyaaba, Shi, et al. (2024) surveyed Ghana teachers in using Generative AI as a teaching assistant. The teachers had not yet systematically applied GenAI tools in classroom settings, but had leveraged GenAI for various instructional activities, such as identifying assessment strategies (see Figure 3). However, these explorations were not yet consistent or fully integrated into the curriculum or accompanied by specific pedagogy. Instead, teachers were in a phase of trial and refinement, testing GenAI applications to see what works best for their students and teaching style. The literature indicates that this exploratory phase is critical for building teachers’ confidence and competence in using GenAI, setting the stage for more sustained and effective integration.

As teachers start to recognize the tangible benefits of GenAI in education, they see GenAI not just as a theoretical concept but as a practical tool that can enhance student engagement, personalize learning experiences, and streamline administrative tasks. This evolving perception is often fueled by initial successes in using GenAI tools, which reinforce the idea that GenAI can complement rather than replace their teaching practices.

As teachers gain hands-on experience with GenAI tools, their initial skepticism diminishes, replaced by a cautious optimism about the technology’s potential. This growing acceptance is crucial for sustained GenAI integration, as it reflects an increasing confidence in the technology’s ability to enhance educational outcomes. Teachers at this stage are more open to exploring new AI tools and are willing to invest time in learning how to use them effectively.

Teachers’ knowledge of GenAI also deepens during this stage, moving beyond basic theoretical understanding to acquire practical knowledge about GenAI concepts such as machine learning, natural language processing, and their specific applications in educational tools. This knowledge is often gained through professional development programs, online courses, or collaborative learning with peers. Teachers become more adept at navigating GenAI interfaces, understanding how these systems generate content, and identifying ways to incorporate GenAI-driven tools into their lessons.

\begin{figure}[bt]
\centering
\includegraphics[width=12cm]{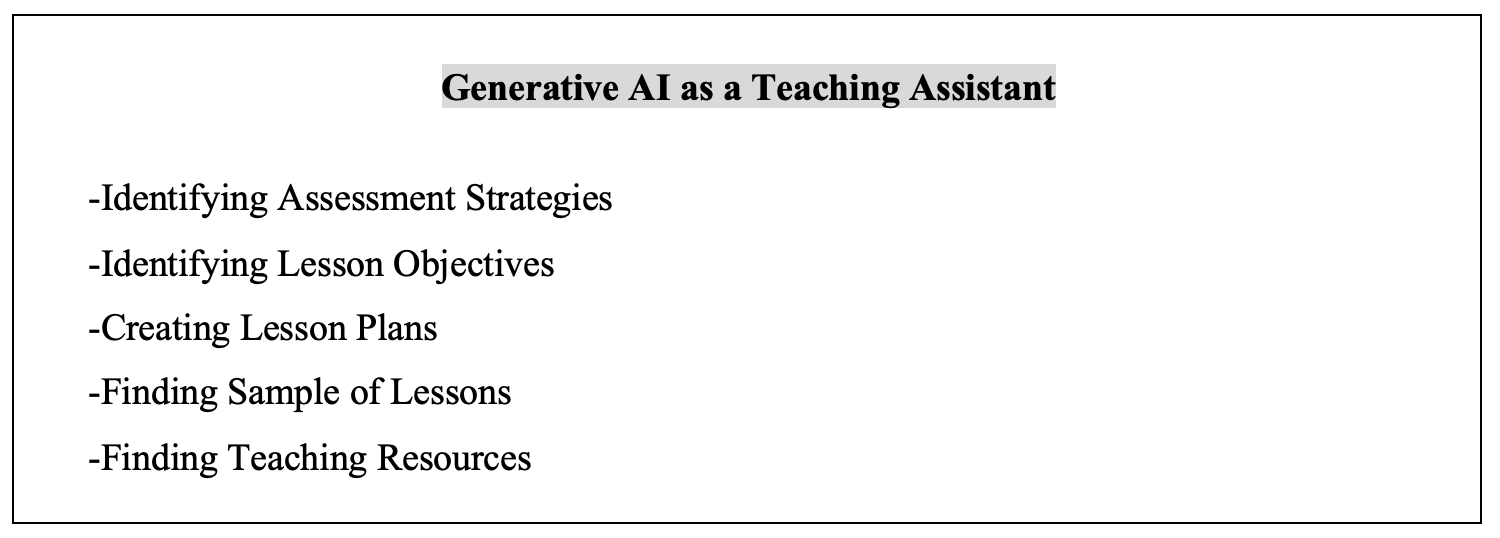}
\caption{Activities that Teachers Reported Using GenAI as a Teaching Assistant (Nyaaba, Shi, et al., 2024)}
\end{figure}

\subsection{Integration Stage: GenAI Collaborator}
In the Integration stage, teachers move beyond exploration to fully incorporate GenAI into their teaching practices. This stage is characterized by seamless integration of AI tools into the educational process, where teachers use GenAI consistently and purposefully to enhance instructional outcomes and student learning experiences. The literature on educational technology integration highlights this stage as one where GenAI becomes an indispensable part of the teaching practice, with teachers taking GenAI as a collaborator and leveraging GenAI’s capabilities to transform their pedagogical approaches.

At the Integration stage, teachers use GenAI tools consistently and purposefully as an integral part of their pedagogy, employing them to personalize learning experiences, automate administrative tasks, and support student assessment. For example, Lee and Zhai (2024b) reported that an instructor engaged student teachers to incorporate ChatGPT in lesson planning. The instructor observed that student teachers incorporated ChatGPT in planned learning activities in various ways (see Figure 4). In this example, the GenAI technology is used not just as a supplement but as a core component of the educational process. Teachers at this stage have developed routines and strategies for effectively incorporating GenAI into their lessons, resulting in enhanced student outcomes and more efficient teaching practices. 

At the Integration stage, teachers’ perceptions reflect a mature understanding of the role of GenAI in education. Teachers perceive GenAI as a valuable and integral component of modern education, recognizing its potential to transform both teaching and learning. They understand that GenAI can not only enhance the efficiency and effectiveness of instruction but also open up new possibilities for personalized learning, student engagement, and formative assessment. This positive perception is grounded in experience, as teachers have seen firsthand the benefits that GenAI can bring to their classrooms.

Teachers are confident in their ability to use GenAI tools effectively and are committed to continuous learning about GenAI to keep pace with technological advancements. The literature suggests that this high level of acceptance is associated with a proactive attitude towards professional development, where teachers seek out opportunities to deepen their GenAI expertise and explore new GenAI applications. This stage also sees teachers becoming advocates for GenAI integration, sharing their experiences and insights with colleagues and contributing to a broader culture of GenAI adoption within their schools (Nyaaba, Zhai, et al., 2024).

Teachers possess intermediate to advanced knowledge of GenAI--a comprehensive understanding of AI’s practical applications, ethical considerations, and potential challenges. They are capable of critically evaluating AI tools for classroom use, ensuring that these tools align with their pedagogical goals and meet the needs of their students. This advanced knowledge allows teachers to make informed decisions about how to best integrate AI into their teaching, balancing the use of AI-generated content with human oversight and creativity.
\begin{figure}[bt]
\centering
\includegraphics[width=12cm]{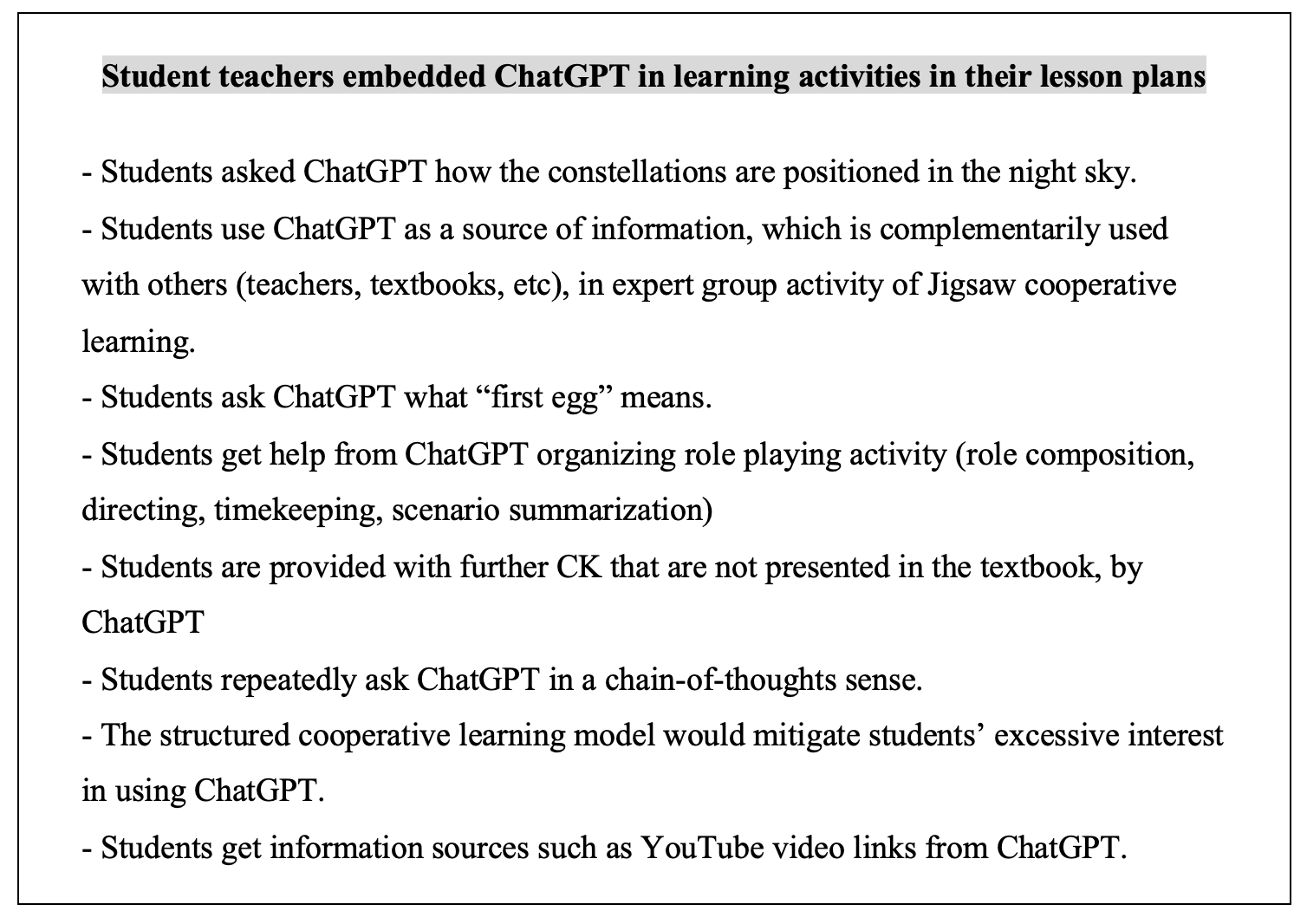}
\caption{Example of incorporating ChatGPT in lesson plans in a teaching method class (Lee \& Zhai, 2024b)}
\end{figure}

\subsection{Innovation Stage: GenAI Innovator}

The final stage in the framework, the Innovation stage, is where teachers become pioneers in the use of GenAI, pushing the boundaries of how the technology can be applied in education. At this stage, teachers are not just users of GenAI tools but innovators and thought leaders, actively shaping the future of GenAI in education. Research on educational innovation highlights the significance of this stage, where teachers contribute to the development of new GenAI-driven educational paradigms and inspire others to explore the transformative potential of GenAI.

At the Innovation stage, teachers usually show substantial leadership and creativity, leading initiatives to transform teaching and learning through GenAI, developing GenAI-based curricula, mentoring other educators, and collaborating on GenAI-related research projects. They may design and implement innovative GenAI-driven educational strategies that challenge traditional pedagogical models and offer new ways of thinking about teaching and learning. For example, Lee, Mun, et al. (2023) designed and developed GenAI application, Collaborative Learning with AI Speakers (CLAIS), using NUGU AI speaker platform. To engage learners in GenAI-involved collaborative learning, they developed a Jigsaw pedagogy—Students were first grouped, and then sent to an expert group with CLAIS specifically focusing on one area of learning. In this expert group, students can communicate with each other, as well as interact with the CLAIS to gain learning support. Afterward, students return to their home group to share what they learned with group members (see Figure 5). In this process, teachers innovatively designed and incorporated GenAI in collaborative learning, thus transforming learning processes. The literature underscores the importance of this stage for the broader adoption of GenAI in education, as these innovative practices serve as models and inspiration for other educators to follow.

Teachers’ perceptions of GenAI are forward-thinking and visionary-- seeing GenAI as a critical driver of innovation in education, with the potential to create entirely new teaching and learning paradigms. They recognize that GenAI is not just a tool to enhance existing practices but a catalyst for reimagining what education can be in the digital age. This perception is informed by a deep understanding of AI’s capabilities and a commitment to exploring its full potential, and can significantly improve teachers’ productivity (Yin et al., 2024). 

At this stage, teachers usually possess advanced and specialized knowledge of GenAI, indicating a deep and nuanced understanding of GenAI’s latest developments, emerging trends, and future potential. They are familiar with cutting-edge GenAI applications and may even contribute to the development of new GenAI applications in education. This advanced knowledge allows them to engage with GenAI at a level that goes beyond practical use, involving them in the design, testing, and refinement of GenAI-driven educational strategies.

Not surprisingly, teachers’ acceptance of GenAI at this stage is complete and enthusiastic, and they not only advocate for AI in education but also lead its adoption and promotion. They actively seek out new AI technologies, experiment with emerging trends, and share their findings with the broader educational community. The literature suggests that teachers at this stage play a crucial role in driving systemic change, influencing both policy and practice through their pioneering work with GenAI.

\begin{figure}[bt]
\centering
\includegraphics[width=11cm]{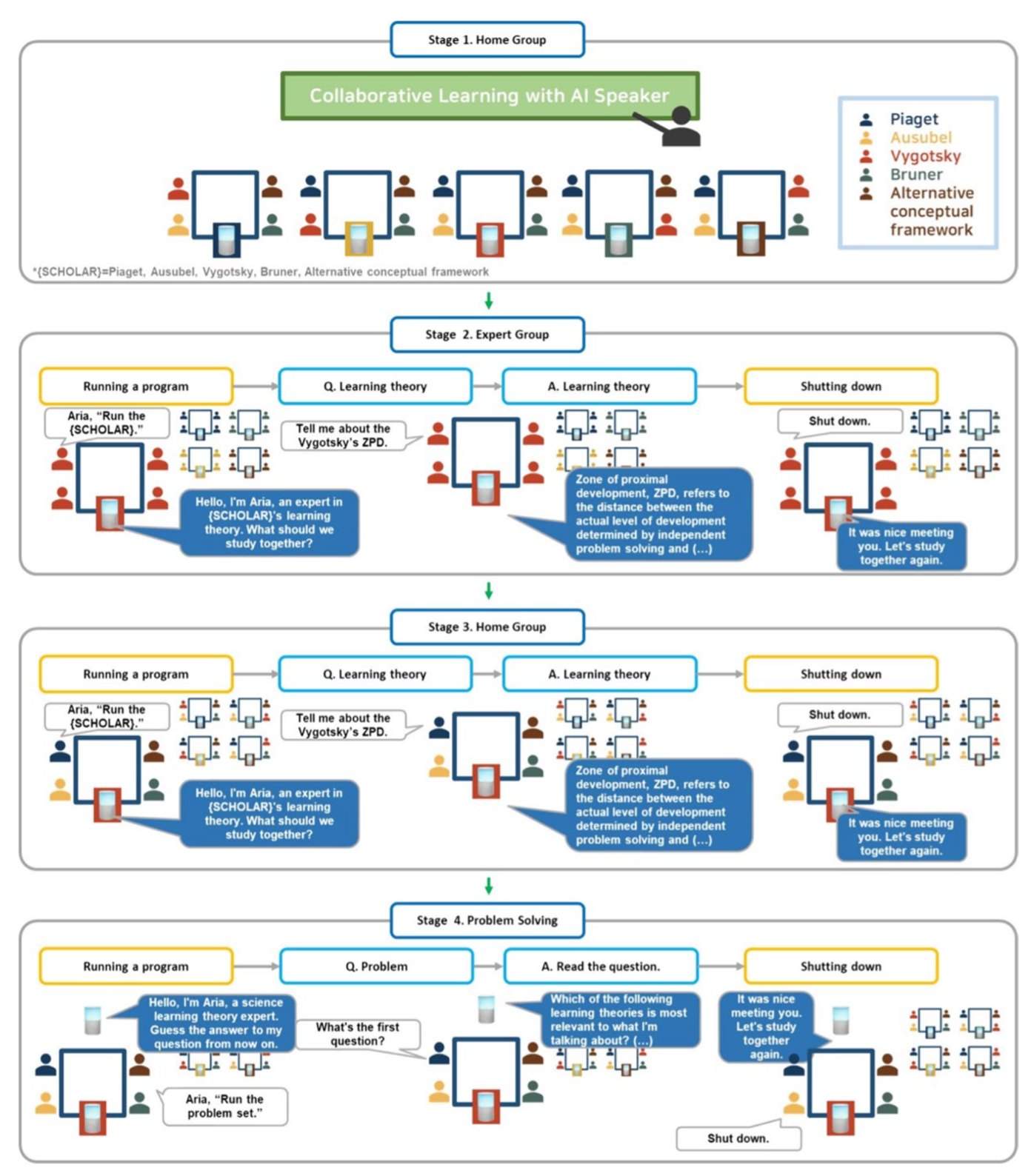}
\caption{The practice of collaborative learning with CLAIS (Lee, Mun, et al., 2023)}
\end{figure}

\section{Implications and Conclusions}

This paper contributes to the literature by highlighting the changes to teachers’ roles and agencies in the era of GenAI and considering teachers’ perceptions, acceptance, knowledge, and practices of GenAI as interconnected components that must be addressed holistically. Current research often isolates these factors, overlooking how they collectively influence teachers’ ability to effectively integrate GenAI into their classrooms. By examining these dimensions in concert, we provide a more comprehensive understanding of how teachers engage with GenAI, shedding light on the complexities of adoption and the varying levels of integration as teachers alter their roles and agencies in classroom settings. This nuanced approach allows for more targeted strategies to enhance teachers’ GenAI competencies, which is crucial for successful educational reform in this digital era.

Furthermore, the framework underscores the importance of viewing GenAI as more than just a tool for enhancing existing practices. It has the potential to transform educational practices at a fundamental level (Lee \& Zhai, 2024b). Teachers’ evolving roles and agencies in the era of GenAI—from observers and adopters to collaborators and innovators—require continuous professional development and institutional support. This paper demonstrates that without a solid foundation of perceptions, knowledge, and acceptance, teachers are less likely to move beyond the superficial uses of GenAI. Thus, it is imperative to foster an environment where teachers can experiment, collaborate, and innovate with GenAI to fully realize its pedagogical potential.

In addition, this study highlights the necessity of supporting teachers in not only integrating GenAI but also in evolving their roles and agencies to deeply embed the technology into their instructional practices. The study stresses that teachers who receive comprehensive training and support are more likely to engage with GenAI meaningfully, moving beyond basic uses to employ it as a tool for personalized learning, adaptive feedback, and co-creative educational experiences. This transformation can only occur if educators are equipped with the skills and knowledge to critically assess and apply GenAI, ensuring that it enhances the learning environment while addressing challenges such as bias and ethical concerns (Lee, Mun, et al., 2023).

In conclusion, this paper provides a comprehensive framework for understanding the evolving roles and agencies of teachers in the era of GenAI, emphasizing the importance of addressing their perceptions, knowledge, acceptance, and practices as interconnected components. By moving beyond a fragmented view of these factors, we offer insights into how educators can fully integrate GenAI into their pedagogical practices, transforming their roles and agencies from passive users to active innovators. The framework underscores the need for ongoing professional development and institutional support to ensure teachers can effectively harness the potential of GenAI, not only to enhance educational practices but to fundamentally reshape them. As education continues to be reshaped by emerging technologies, the ability of teachers to engage critically and creatively with GenAI will be crucial in driving meaningful, equitable, and impactful learning experiences.

\section*{References}

\begin{hangparas}{.5in}{1}

Ali, F., Choy, D., Divaharan, S., Tay, H. Y., \& Chen, W. (2023). Supporting self-directed learning and self-assessment using TeacherGAIA, a generative AI chatbot application: Learning approaches and prompt engineering. Learning: Research and Practice, 9(2), 135-147.

Bewersdorff, A., Hartmann, C., Hornberger, M., Seßler, K., Bannert, M., Kasneci, E., Kasneci, G., Zhai, X., \& Nerdel, C. (2024, 2024/01/01/). Taking the Next Step with Generative Artificial Intelligence: The Transformative Role of Multimodal Large Language Models in Science Education. arXiv preprint arXiv:2401.00832. 

Bewersdorff, A., Zhai, X., Roberts, J., \& Nerdel, C. (2023). Myths, mis- and preconceptions of artificial intelligence: A review of the literature. Computers and Education: Artificial Intelligence, 4(100143), 1-11. \url{https://doi.org/10.1016/j.caeai.2023.100143} 

Celik, I. (2023). Towards Intelligent-TPACK: An empirical study on teachers’ professional knowledge to ethically integrate artificial intelligence (AI)-based tools into education. Computers in Human Behavior, 138. \url{https://doi.org/10.1016/j.chb.2022.107468}

Collie, R. J., \& Martin, A. J. (2024). Teachers’ motivation and engagement to harness generative AI for teaching and learning: The role of contextual, occupational, and background factors. Computers and Education: Artificial Intelligence, 6, 100224. 

Cotton, D. R., Cotton, P. A., \& Shipway, J. R. (2024). Chatting and cheating: Ensuring academic integrity in the era of ChatGPT. Innovations in Education and Teaching International, 61(2), 228-239. 

Davis, F. D., Bagozzi, R., \& Warshaw, P. (1989). Technology acceptance model. J Manag Sci, 35(8), 982-1003. 

Feldman-Maggor, Y., Blonder, R., \& Alexandron, G. (2024). Perspectives of Generative AI in Chemistry Education Within the TPACK Framework. Journal of Science Education and Technology, 1-12. 

Garofalo, S. G., \& Farenga, S. J. (2024). Science Teacher Perceptions of the State of Knowledge and Education at the Advent of Generative Artificial Intelligence Popularity. Science \& Education, 1-20. 

Goldberg, B., \& Robson, R. (2023). AI to Support Guided Experiential Learning. International Conference on Artificial Intelligence in Education, 

Guo, S., Shi, L., \& Zhai, X. (2024). Validating an Instrument for Teachers' Acceptance of Artificial Intelligence in Education. arXiv preprint arXiv:2406.10506. 

Jo, A. (2023). The promise and peril of generative AI. Nature, 614(1), 214-216. 

Kaplan-Rakowski, R., Grotewold, K., Hartwick, P., \& Papin, K. (2023). Generative AI and teachers’ perspectives on its implementation in education. Journal of Interactive Learning Research, 34(2), 313-338. 

Kizilcec, R. F., Huber, E., Papanastasiou, E. C., Cram, A., Makridis, C. A., Smolansky, A., Zeivots, S., \& Raduescu, C. (2024). Perceived impact of generative AI on assessments: Comparing educator and student perspectives in Australia, Cyprus, and the United States. Computers and Education: Artificial Intelligence, 7, 100269. 

Kong, S.-C., \& Yang, Y. (2024). A Human-Centred Learning and Teaching Framework Using Generative Artificial Intelligence for Self-Regulated Learning Development through Domain Knowledge Learning in K–12 Settings. IEEE Transactions on Learning Technologies. 

Laak, K.-J., \& Aru, J. (2024). Generative AI in K-12: Opportunities for Learning and Utility for Teachers. International Conference on Artificial Intelligence in Education, 

Lan, Y.-J., \& Chen, N.-S. (2024). Teachers’ agency in the era of LLM and generative AI. Educational Technology \& Society, 27(1), I-XVIII. 

Latif, E., Parasuraman, R., \& Zhai, X. (2024). PhysicsAssistant: An LLM-Powered Interactive Learning Robot for Physics Lab Investigations. IEEE International Conference on Robotics and Automation (ICRA), 

Lee, D., Arnold, M., Srivastava, A., Plastow, K., Strelan, P., Ploeckl, F., Lekkas, D., \& Palmer, E. (2024). The impact of generative AI on higher education learning and teaching: A study of educators’ perspectives. Computers and Education: Artificial Intelligence, 6, 100221. 

Lee, G.-G., Latif, E., Shi, L., \& Zhai, X. (2023). Gemini Pro Defeated by GPT-4V: Evidence from Education. arXiv preprint arXiv:2401.08660. 

Lee, G.-G., Latif, E., Wu, X., Liu, N., \& Zhai, X. (2024, 2023/11/30/). Applying Large Language Models and Chain-of-Thought for Automatic Scoring. Computers \& Education: Artificial Intellligence, 6(100213), 1-15. \url{https://doi.org/https://doi.org/10.1016/j.caeai.2024.100213}

Lee, G.-G., Mun, S., Shin, M.-K., \& Zhai, X. (2023). Collaborative Learning with Artificial Intelligence Speakers (CLAIS): Pre-Service Elementary Science Teachers' Responses to the Prototype. Science \& Education, 1-29. 

Lee, G.-G., \& Zhai, X. (2023). NERIF: GPT-4V for Automatic Scoring of Drawn Models. arXiv preprint arXiv:2311.12990. 

Lee, G.-G., \& Zhai, X. (2024a). Realizing Visual Question Answering for Education: GPT-4V as a Multimodal AI. arXiv preprint arXiv:2405.07163. 

Lee, G.-G., \& Zhai, X. (2024b). Using ChatGPT for science learning: A study on pre-service teachers' lesson planning. IEEE Transactions on Learning Technologies, 17, 1683 - 1700. \url{https://doi.org/10.1109/TLT.2024.3401457} 

Lee, S., \& Song, K.-s. (2024). Teachers' and Students' Perceptions of AI-Generated Concept Explanations: Implications for Integrating Generative AI in Computer Science Education. Computers and Education: Artificial Intelligence, 100283. 

Martin, P. P., \& Graulich, N. (2024). Beyond Language Barriers: Allowing Multiple Languages in Postsecondary Chemistry Classes Through Multilingual Machine Learning. Journal of Science Education and Technology, 1-16. 

Mishra, P., Oster, N., \& Henriksen, D. (2024). Generative AI, Teacher Knowledge and Educational Research: Bridging Short-and Long-Term Perspectives. TechTrends, 68(2), 205-210. 

Mishra, P., Warr, M., \& Islam, R. (2023). TPACK in the age of ChatGPT and Generative AI. Journal of Digital Learning in Teacher Education, 39(4), 235-251. 

Nyaaba, M., Shi, L., Nabang, M., Zhai, X., Kyeremeh, P., Ayoberd, S. A., \& Akanzire, B. N. (2024). Generative AI as a Learning Buddy and Teaching Assistant: Pre-service Teachers' Uses and Attitudes. arXiv preprint arXiv:2407.11983. 

Nyaaba, M., \& Zhai, X. (2024). Generative AI professional development needs for teacher educators. Journal of AI, 8(1), 1-13. 

Nyaaba, M., Zhai, X., \& Faison, M. Z. (2024). Generative AI for Culturally Responsive Assessment in Science: A Conceptual Framework. 

Pahi, K., Hawlader, S., Hicks, E., Zaman, A., \& Phan, V. (2024). Enhancing active learning through collaboration between human teachers and generative AI. Computers and Education Open, 6, 100183. 

Rosenberg, J. M., \& Koehler, M. J. (2015). Context and technological pedagogical content knowledge (TPACK): A systematic review. Journal of Research on Technology in Education, 47(3), 186-210. 

Shi, L., \& Choi, I. (2024). A Systematic Review on Artificial Intelligence in Supporting Teaching Practice: Application Types, Pedagogical Roles, and Technological Characteristics. In X. Zhai \& J. Krajcik (Eds.), Uses of Artificial Intelligence in STEM Education (pp. 321-347). Oxford University Press. 

Srinivasan, R., \& Uchino, K. (2021). Quantifying Confounding Bias in Generative Art: A Case Study. arXiv preprint arXiv:2102.11957. 

Tang, K.-S., \& Cooper, G. (2024). The role of materiality in an era of generative artificial intelligence. Science \& Education, 1-16. 

Tang, K.-S., Cooper, G., Rappa, N., Cooper, M., Sims, C., \& Nonis, K. (2024). A Dialogic Approach to Transform Teaching, Learning \& Assessment with Generative AI in Secondary Education. Learning \& Assessment with Generative AI in Secondary Education (February 11, 2024). 

van den Berg, G., \& du Plessis, E. (2023). ChatGPT and generative AI: Possibilities for its contribution to lesson planning, critical thinking and openness in teacher education. Education Sciences, 13(10), 998. 

Venkatesh, V., Thong, J. Y., \& Xu, X. (2016). Unified theory of acceptance and use of technology: A synthesis and the road ahead. Journal of the association for Information Systems, 17(5), 328-376. 

Yin, H., Wang, C., \& Liu, Z. (2024). Unleashing Pre-Service Language Teachers’ Productivity with Generative AI: Emotions, Appraisal and Coping Strategies. Computers in Human Behavior, 108417. 

Zhai, X. (2023). ChatGPT for Next Generation Science Learning. XRDS: Crossroads. \url{https://doi.org/https://doi.org/10.1145/3589649} 

Zhai, X., \& Krajcik, J. (2024). Pseudo AI Bias. In Uses of Artificial Intelligence in STEM Education. Oxford University Press. \url{https://doi.org/10.48550/arXiv.2210.08141}

Zhai, X., \& Nehm, R. (2023). AI and formative assessment: The train has left the station. Journal of Research in Science Teaching, 60(6), 1390-1398. \url{https://doi.org/DOI: 10.1002/tea.21885}

Zhai, X., Yin, Y., Pellegrino, J. W., Haudek, K. C., \& Shi, L. (2020). Applying machine learning in science assessment: a systematic review. Studies in Science Education, 56(1), 111-151. 

Zhou, M., Abhishek, V., Derdenger, T., Kim, J., \& Srinivasan, K. (2024). Bias in Generative AI. \url{arXiv preprint arXiv:2403.02726.} 

\end{hangparas}

\end{document}